# Supersonic triple deck flow past a flat plate with an elastic stretch

TAREK M. A. EL-MISTIKAWY

Department of Engineering Mathematics and Physics, Faculty of Engineering, Cairo University, Giza 12211, Egypt.

**Abstract**
The steady laminar supersonic flow past a flat plate having a stretch of an elastic membrane, the pressure on the other side of which is adjustable, is studied within the framework of the triple deck theory. The resulting lower deck problem is supplemented with a membrane equation relating the pressure difference across the membrane to its curvature. By pressurizing or depressurizing the membrane, it assumes the form of a hump or a dent that alters the flow characteristics. Numerical solutions obtained, in either case, give plausible account of the interaction between the membrane and the flow.

## 1. INTRODUCTION

The triple deck theory [1] establishes a matched asymptotic structure that explains how a boundary layer flow senses a downstream disturbance and interacts with the outer inviscid flow to facilitate non-singular development past the disturbance. The theory proved successful in handling the analysis problem of finding the flow characteristics, given the body shape or source of disturbance; even in cases involving flow separation [2].

The triple deck analysis leads to a lower deck problem that has to be solved numerically. The governing equations are identical to those of the incompressible boundary layer equations; however, with a complication. The pressure is no longer specified, but needs to be determined in the course of the solution to satisfy a viscid/inviscid interaction law.

A further complication can be encountered, when the surface shape, as well, needs to be determined in the course of the solution, to satisfy either a design requirement or a moving boundary condition. The first design problem was introduced by El-Mistikawy [3] who designed a compression corner with prescribed minimum surface shear. El-Mistikawy and El-Fayez treated the first moving boundary problem in their study of the flow past an eroding hump [4]. Lagrée [5] applied a quasi-steady approach to study ripple formation and evolution.

Another interesting problem, that is the subject of the present study, is the flow past a surface having a stretch of an elastic membrane; the pressure on the other side of which is adjustable. By varying this pressure, the membrane assumes a hump or a dent shape that alters the flow characteristics. The lower deck problem needs to be solved in conjunction with a membrane equation relating the pressure difference across the membrane to its curvature; so that the flow field and the shape of the membrane can be determined. The problem is solved by El-Mistikawy's [3] supersonic solver, which is modified to account for this new feature. The membrane shape is approximated by cubic splines to guarantee smoothness.

Related problems that can be found in the literature are flows past solid surfaces that contain humps or dents. The hump problem was first formulated by Smith [6] and solved numerically by Napolitano, Davis and Werle [7]. For moderate hump heights, a separation bubble forms at the trailing edge of the hump. Another bubble is expected to form at the leading edge, but for much higher humps. For the dent problem, interacting boundary layer models, based on the triple deck theory, were constructed and solved [8,9]. They reveal that a separation bubble forms at the bottom of the dent.

Solutions of the present problem are obtained for cases of pressurized and depressurized membranes. A pressurized membrane curves inward (i.e. toward the flow) as a hump, whose convex shape causes the flow to expand; reducing its pressure. A depressurized membrane, on the other hand, curves outward as a dent, whose concave shape causes the flow to be compressed. In either case, there results an increase in the pressure difference across the membrane, and consequently in the membrane deformation. However, this effect is more pronounced in the former case. It is also observed that the hump, being exposed to the flow, becomes asymmetric- leaning toward the lie side; whereas the dent, being shunned from the flow, assumes an almost symmetric shape. As with the solid configurations, separation bubbles form at the trailing edge of a hump and at the bottom of a dent, when the membrane is sufficiently deformed.

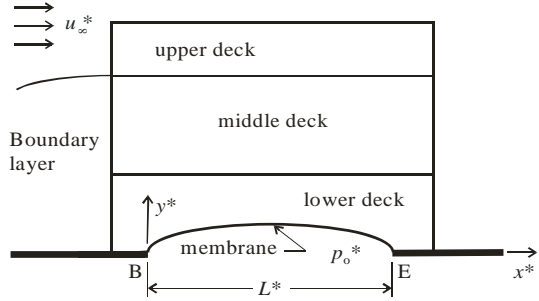

Fig. 1: Flow Configuration

## 2. STATEMENT OF THE PROBLEM

A fully developed two-dimensional laminar supersonic boundary layer flow passes above a flat plate having a stretch BE of length $L^*$ of an elastic membrane, as shown in Fig. 1. At B, we introduce Cartesian coordinate axes: $x^*$ along the plate pointing downstream, and $y^*$ normal to the plate. The pressure $p_o^*$ below the membrane is adjustable; causing the membrane to curve into a hump or a dent that is described by $y^*=f^*(x^*)$ for $0 \leq x^* \leq L^*$; where for a hump $f^*>0$, while for a dent $f^*<0$. (Henceforth, $f^*$ is extended to describe the entire surface, given zero value outside the interval $0 \leq x^* \leq L^*$.)

Far from the surface, the freestream has uniform velocity $u_\infty^*$ in the $x^*$ direction, density $\rho_\infty^*$, pressure $p_\infty^*$, sonic speed $a_\infty^*$, and viscosity $\mu_\infty^*$. The freestream is supersonic with Mach number $Ma=u_\infty^*/a_\infty^*>1$. The Reynolds number $Re=\rho_\infty^* u_\infty^* L^*/\mu_\infty^*$ is large and relates to a diminishing parameter $\varepsilon=Re^{-1/5}$.

For $f^*=O(\varepsilon^2 L^*)$, the boundary layer structure approaching the membrane develops into a triple deck structure containing the membrane within its lower deck, which is governed by the following problem [6]

$U_{,X}+V_{,Y}=0$ (1)
$U U_{,X}+U_{,Y}V+P_{,X}=U_{,YY}$ , $P_{,Y}=0$ (2a,b)
$U=0$ , $V=0$ at $Y=F(X)$ (3a,b)
$U_{,Y} \sim 1$ , $U \sim Y-A$ as $Y \sim \infty$ (4a,b)
$U \sim Y$ , $A \sim 0$ as $X \sim -\infty$ (5a,b)
$A \sim 0$ as $X \sim \infty$ (6)
$P = A_{,X}$ (7)

where subscripts following a comma denote differentiation, and $A(X)$ is the displacement function.

The normalized variables: Cartesian coordinates $(X,Y)$, membrane height $F$, velocity components $(U,V)$, and pressure $P$ are related to the corresponding physical variables $x^*, y^*, f^*, u^*, v^*$, and $p^*$ as follows

$x^*=\rho_s^{-1/2}\mu_s^{-1/4}\beta^{-3/4}\tau^{5/4} X L^*$ (8a)
$(y^*,f^*)=\varepsilon^2 \rho_s^{-1/2}\mu_s^{1/4}\beta^{-1/4}\tau^{3/4} (Y,F) L^*$ (8b)
$u^*=\varepsilon \rho_s^{-1/2}\mu_s^{1/4}\beta^{-1/4}\tau^{-1/4} U u_\infty^*$ (8c)
$v^*=\varepsilon^3 \rho_s^{-1/2}\mu_s^{3/4}\beta^{1/4}\tau^{-3/4} V u_\infty^*$ (8d)
$p^*-p_\infty^*=\varepsilon^2 \mu_s^{1/2}\beta^{-1/2}\tau^{1/2} P \rho_\infty^* u_\infty^{*2}$ (8e)

where the subscript s denotes values at the surface, $\rho_s = \rho_s^*/\rho_\infty^*$, $\mu_s = \mu_s^*/\mu_\infty^*$, and $\beta = (Ma^2-1)^{1/2}$ is the Mach factor; while $\tau$ represents the surface shear of the oncoming boundary layer, as it approaches the hump region.

The lower deck problem is supplemented with the membrane equation derived, as usual, by considering the equilibrium of a membrane element of length $dl^*$ under the action of the membrane tension forces at its ends and the force due to the pressure difference across the element (See Fig. 2.). In dimensional variables, this yields

$$p^* - p_o^* = T^* d\theta/dl^* \tag{9a}$$

where $\theta$ is the angle of inclination to the $x^*$ axis; and $T^*$ is the membrane tension whose streamwise variation, due to the shearing action of the flow, is neglected- being $O(\varepsilon^4)$. Substituting $\theta \approx df^*/dx^*$ and $dx^*/dl^* \approx 1$, with $O(\varepsilon^4)$ errors, gives

$$p^* - p_o^* = T^* d^2 f^*/dx^{*2} \tag{9b}$$

In normalized form, with

$$T^* = \rho_s^{-1/2} \mu_s^{-1/4} \beta^{-7/4} \tau^{-9/4} \sigma \rho_\infty^* u_\infty^{*2} L^* \tag{9c}$$

the membrane equation takes the form

$$P - P_o = \sigma F_{,XX} \tag{10}$$

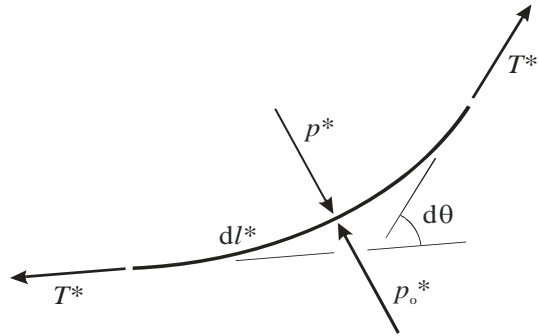

Fig. 2: Forces acting on a membrane element

## 3. NUMERICAL SOLUTION

The lower deck problem (1-7) and the membrane equation (10) are simultaneously solved for the flow variables and the membrane shape. Prandtl's shift [10]

$$y = Y - F \ , \ v = V - U F_{,X} \tag{11a,b}$$

is applied to transform the membrane surface to the line $y=0$. So shifted, the problem takes the new form

$$U_{,X} + v_{,y} = 0 \tag{12}$$
$$U U_{,X} + U_{,y} v + Q = W_{,y} \ , \ W = U_{,y} \tag{13a,b}$$
$$U = 0 \ , \ v = 0 \qquad \text{at } y = 0 \tag{14a,b}$$
$$W \sim 1 \ , \ U \sim y - A + F \quad \text{as } y \sim \infty \tag{15a,b}$$
$$U \sim y \ , \ A \sim 0 \qquad \text{as } X \sim -\infty \tag{16a,b}$$

$$A \sim 0 \qquad \text{as } X \sim \infty \tag{17}$$
$$P = A_{,X}, \quad Q = P_{,X} \tag{18a,b}$$
$$P - P_o = \sigma F_{,XX} \tag{19}$$

where the shear function $W(X,y)$ and the pressure gradient function $Q(X)$ are defined by Eqs. (13b) and (18b), respectively.

### 3.1. Iterative Procedure

The problem is solved by the supersonic triple-deck solver of El-Mistikawy [3]. It is an efficient iterative procedure that takes, in the $k^{th}$ iteration, an old distribution $A^k$ of the displacement function and produces a new one $A^{k+1}$. All equations and conditions (12-19) are expressed in the new iteration level k+1, except for Condition (15b) which is written as

$$U^{k+1} \sim y - A^k + F^k \qquad \text{as } y \sim \infty \tag{15b'}$$

and Eq. (18b) which is replaced by the relaxation relation

$$P^{k+1}{}_{,X} - Q^{k+1} = r(A^{k+1} - A^k) \tag{18b'}$$

where $r$ is a relaxation factor. As convergence is reached, $A^{k+1} \approx A^k$; then, Eq. (18b') approaches Eq. (18b).

For convenience, the superscript k+1 is dropped, henceforth.

### 3.2. Discretization Layout

The computational domain covers the region of the $Xy$-plane described by $X_{-\infty} \leq X \leq X_{+\infty}$, $0 \leq y \leq y_{+\infty}$, where $-X_{-\infty}$, $X_{+\infty}$ and $y_{+\infty}$ are large enough to allow for adequate enforcement of the asymptotic conditions (15-17). It is divided into rectangular cells by a grid of (i)-lines $X=X(i)$ and (j)-lines $y=y(j)$. The counter i=1→I is such that $X(1)=X_{-\infty}$, $X(i_B)=0$, $X(i_E)=1$, and $X(I)=X_{+\infty}$; while the counter j=1→J is such that $y(1)=0$ and $y(J)=y_{+\infty}$. The typical ij-cell has $X$-step $\Delta(i)=X(i)-X(i-1)$ and $y$-step $\delta(j)=y(j)-y(j-1)$, and through its midpoint pass the (i-½) and (j-½)-lines.

The discretization layout is as follows: $U$ is assigned to the grid points (i,j), whereas $v$ and $W$ are assigned to the points (i-½,j). As for the $X$-dependent variables, $A$, $P$ and F are assigned to the i-lines, whereas $Q$ is assigned to the (i-½)-lines.

### 3.3. Solution Steps

Each iteration involves two sweeps. The first sweep applies streamwise marching to solve for $U$, $v$, $W$, $Q$ and $F$. Equations (12-13) are centered at the midpoint of the ij-cell. Central difference expressions, in terms of the abovementioned layout, are used throughout. However, $U$ and $U_{,y}$ are expressed in terms of $U(i-½,j)$ and $U(i-½,j-1)$, then each $U(i-½,j)$ is expressed backwards as $U(i-1,j)+½\Delta(i)U_{,X}(i-½,j)$ in Eq. (13a) and forwards as $U(i,j)-½\Delta(i)U_{,X}(i-½,j)$ in Eq. (13b). For these representations to be second order accurate in the $X$-direction, $U_{,X}(i-½,j)$ may be represented with first order accuracy. At i=2, we set $U_{,X}(i-½,j)=0$ in accordance with Condition (16a). Next to the two lines i=$i_B$ and i=$i_E$ where $U_{,X}$ is discontinuous, we use the respective $k^{th}$ iterate $U^k{}_{,X}(i+3/2,j)$ with the representation $[U^k(i+2,j)-U^k(i+1,j)]/\Delta(i+2)$ to approximate $U_{,X}(i+½,j)$ and to represent $U_{,X}(i+3/2,j)$. Otherwise, since continuity permits, $U_{,X}(i-½,j)$ is approximated by its counterpart at the (i-3/2)-line with the representation $[U(i-1,j)-U(i-2,j)]/\Delta(i-1)$.

At each i-line, the 3J−3 finite difference equations are supplemented, in view of Conditions (14) and (15), with $U(i,1)=0$, $v(i-½,1)=0$, $W(i-½,J)=1$, and $U(i,J)=y(J)-A^k(i)+F^k(i)$. The resulting set of equations is efficiently solved [11] to yield the 3J+1 unknowns: $U(i,j)$, $v(i-½,j)$, and $W(i-½,j)$ for j=1→J, as well as $Q(i-½)$. Marching starts at i=2, where Conditions (16) supply the needed values at i=1, and ends when i=I.

Next, for $i_B \leq i \leq i_E$, Eq. (19) is used to determine $F_{,XX}(i)$. The resulting $i_E-i_B+1$ values together with the end conditions $F(i_B)=F(i_E)=0$ suffice to determine $F(i)$ for $i_B<i<i_E$, when $F$ is represented by cubic splines so that $F$, $F_{,X}$, and $F_{,XX}$ are continuous at each i-line. The details are given in the Appendix. This marks the end of the first sweep.

In the second sweep, we solve for $P$ and $A$. Equations (18a,b') are centered at the (i-½)-lines and central difference representations are used to achieve second order accuracy. Conditions (16b) and (17) complete the set of equations which are solved to give $P(i)$ and $A(i)$ for $i=1\rightarrow I$.

The method requires very low storage. Of the $k^{th}$ iterates only $A^k(i)$, $F^k(i)$, $U^k_{,X}(i_B+3/2,j)$, and $U^k_{,X}(i_E+3/2,j)$ need to be stored for use in the next iteration.

The iterative process continues until $A$ and $F$ reach convergence as judged by the satisfaction, everywhere, of the inequalities $|A^{k+1}(i)-A^k(i)|<\eta_A$ and $|F^{k+1}(i)-F^k(i)|<\eta_F$ where $\eta_A$ and $\eta_F$ are prescribed tolerances.

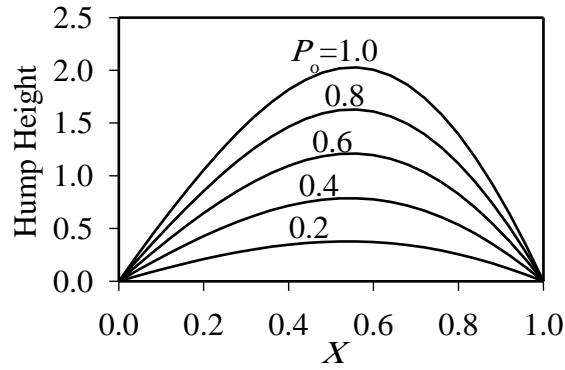

Fig. 3a: Membrane shape ($P_o>0$)

## 4. IMPLEMENTATION

The method was applied on a computational domain extending to $X_{-\infty}=-10$, $X_{+\infty}=16$, and $y_{+\infty}=12$. The grid was such that I=131, J=39, $i_B=36$, and $i_E=61$. Variable step sizes in both directions were utilized to suit regions of fast or slow changes. The X-step size $\Delta$ took the values: 0.4 (for 1<i≤6 and 116<i≤131), 0.3 (for 6<i≤26 and 96<i≤116), 0.2 (for 26<i≤36 and 86<i≤96), 0.04 (for 36<i≤61 and 72<i≤76), 0.05 (for 76<i≤80), and 0.1 (for 80<i≤86). This left a gap of size 0.04 that followed i=$i_E$. It was divided by 10 i-lines into much smaller 11 steps; 5 of size 0.002, 3 of size 0.004, then 3 of size 0.006. The y-step size $\delta$ took the values: 0.02, 0.04, 0.06, and 0.08 each for one step, 0.1 for 2 steps, 0.2 for 6 steps, 0.3 for 8 steps, 0.4 for 10 steps, then 0.5 for the last 8 steps. For convergence tolerances, we chose $\eta_A=\eta_F=10^{-4}$. The first iterates (when k=1) for $A^k$ and $F^k$ were taken to be the parabola $y=4X(1-X)$, while those for $U^k_{,X}(i_B+3/2,j)$ and $U^k_{,X}(i_E+3/2,j)$ were taken to be zero. The suitable value for the relaxation factor $r$ was 2.

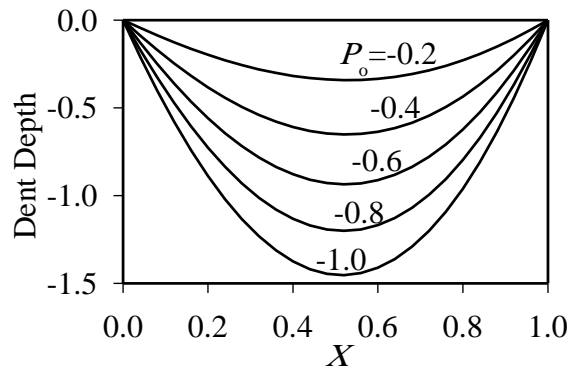

Fig. 3b: Membrane shape ($P_o<0$)

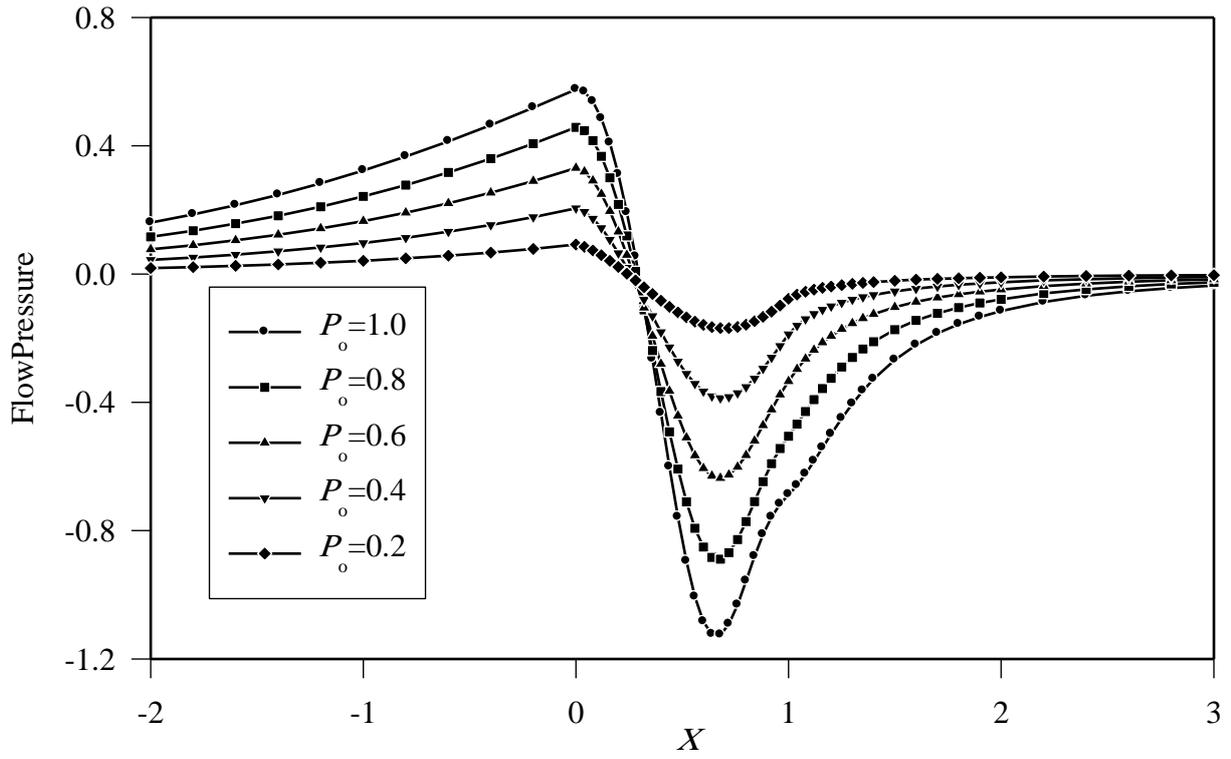

Fig. 4a: Flow Pressure ($P_o>0$)

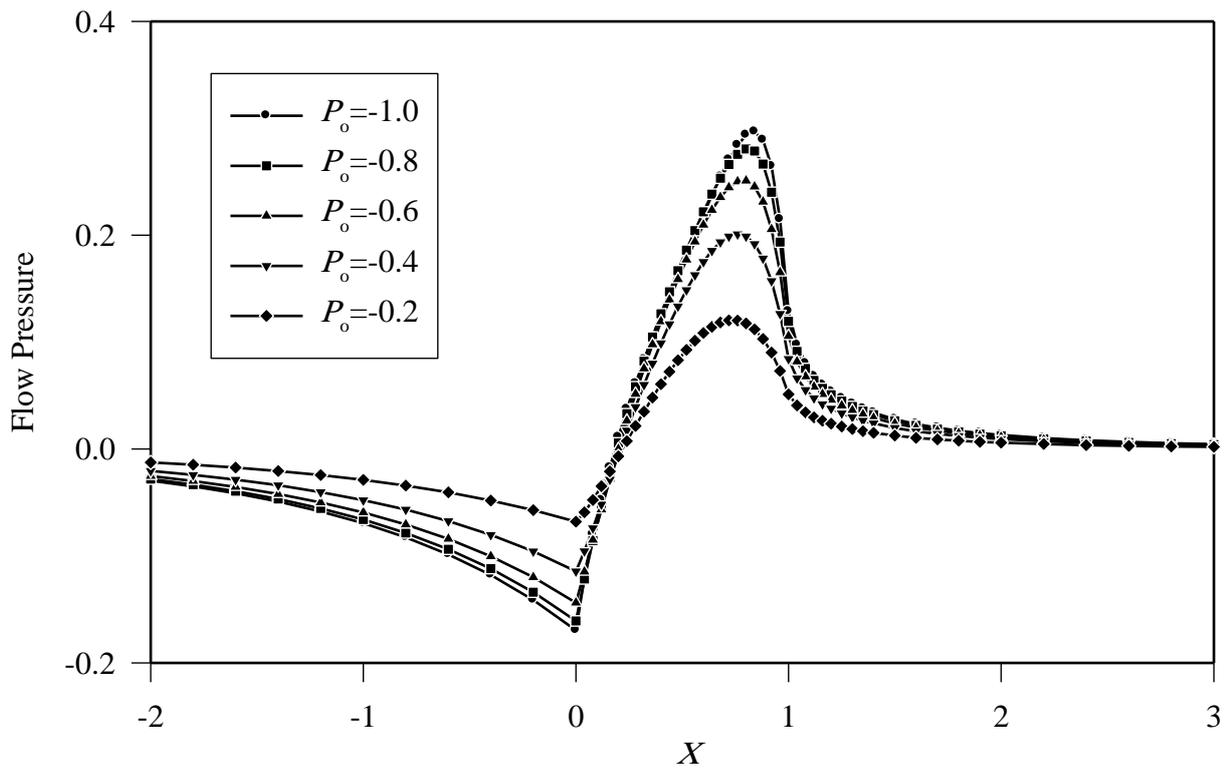

Fig. 4b: Flow Pressure ($P_o<0$)

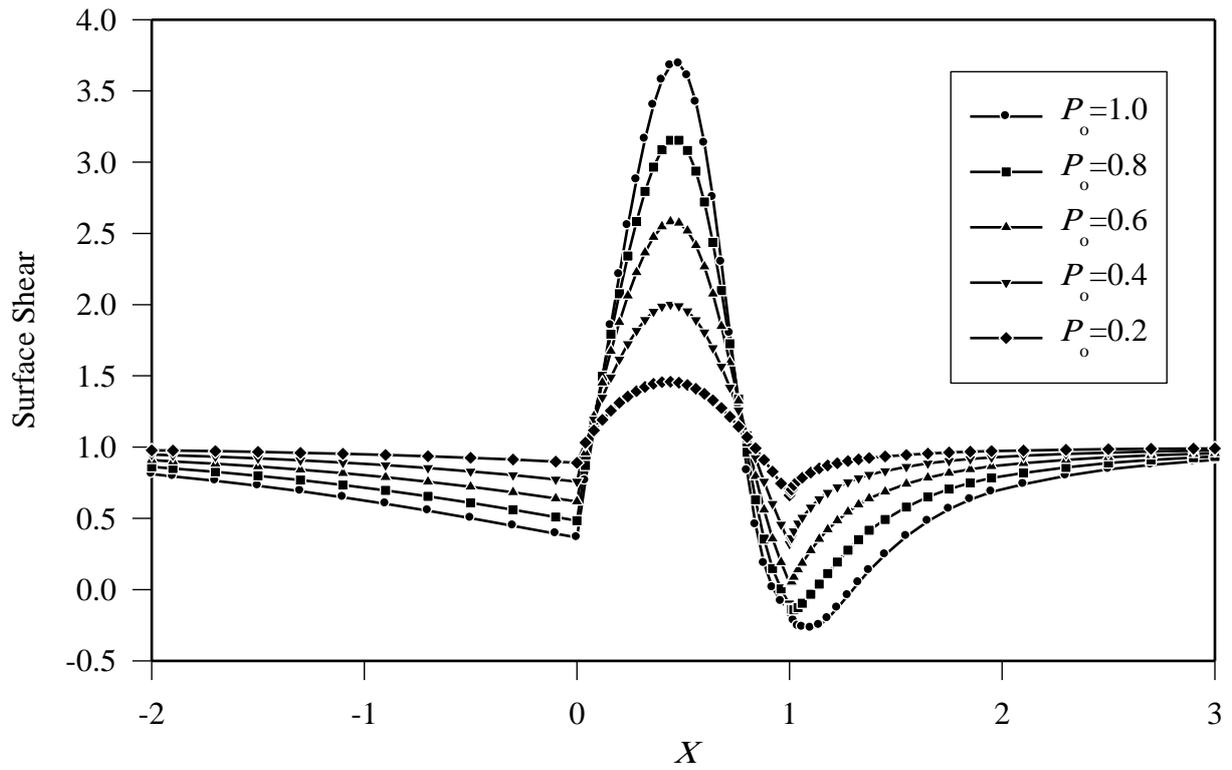

Fig. 5a: Surface Shear ($P_o>0$)

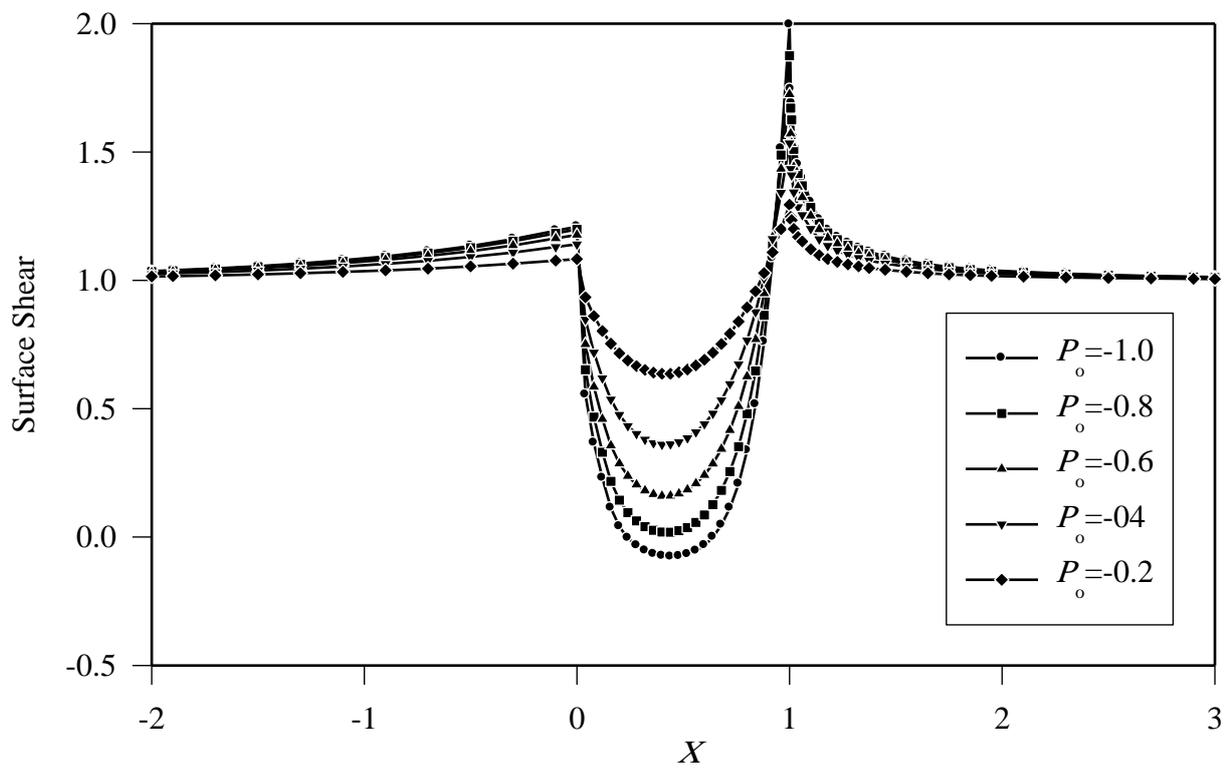

Fig. 5b: Surface Shear ($P_o<0$)

# 5. RESULTS AND DISCUSSION

The problem involves two parameters; the membrane pressure $P_o$ and the membrane tension $\sigma$. The effect of each parameter is studied.

The membrane pressure $P_o$ is varied from $-1$ to $+1$, with $\sigma$ fixed at 0.1. Figures 3a,b show the shape of the membrane for different positive and negative values of $P_o$. As should be expected, the positive values produce humps (Fig. 3a) that are pushed downstream by the flow, to acquire asymmetry; leaning toward the lee side. In contrast, the negative values produce dents (Fig. 3b) that are almost symmetric; being shunned from the flow. The variation with $X$ of the corresponding flow pressure $P(X)$ and surface shear $W(X,0)$ are shown in Figs. 4a,b and 5a,b, respectively.

For positive values of $P_o$, the lower deck pressure (Fig. 4a) rises from the upstream flat plate value of zero to a maximum at the leading edge B of the hump. This first compression stage is a manifestation of the upstream influence detected by the triple deck theory. The pressure, then, drops steeply to a minimum at a point on the lee side; the flow being expanded over the convex shape of the hump. This rather over-expansion is followed by a second compression stage, so that the pressure returns to the flat plate value downstream. The surface shear (Fig. 5a), on the other hand, decreases from the flat plate value of unity to a minimum at B, then rises to a peak at a point on the stern side of the hump. Next, it drops to another minimum but of lower value near the trailing edge E of the hump, before recovering the flat plate value downstream. As $P_o$ is increased, the adverse pressure gradient experienced in the second compression stage becomes strong enough to cause flow separation. The values $P_o=0.8$ and 1.0 exhibit such a behavior as observed in Fig. 5a, in which the surface shear becomes negative near E.

For negative values of $P_o$, the pressure (Fig. 4b) decreases from zero upstream to a minimum at B, then rises to a positive maximum at a point on the lee side close to E, before dropping steeply to zero downstream. The surface shear (Fig. 5b) increases from unity to a maximum at B, then decreases to a minimum just ahead of the midpoint of the membrane, to rise again to a maximum at E, then falls to unity downstream. Large negative values of $P_o$ produce separation bubbles at the bottom of the dent, as observed when $P_o=-1$.

Comparisons of Fig. 4a to Fig. 4b and Fig. 5a to Fig. 5b reveal that the flow is more sensitive to variations in $P_o$, when $P_o$ is positive than when $P_o$ is negative. For the same $|P_o|$, the maxima and minima of the flow pressure and surface shear are much larger when $P_o$ is positive.

Table 1: Effect of membrane tension $\sigma$

| $\sigma$ | Largest Height | | Largest Depth | |
|---|---|---|---|---|
| | value | $X$-place | value | $X$-place |
| 0.10 | 2.0284 | 0.5527 | 1.4535 | 0.5163 |
| 0.12 | 1.5464 | 0.5415 | 1.2098 | 0.5160 |
| 0.14 | 1.2375 | 0.5338 | 1.0313 | 0.5152 |
| 0.16 | 1.0289 | 0.5284 | 0.8958 | 0.5144 |
| 0.18 | 0.8802 | 0.5244 | 0.7901 | 0.5136 |
| 0.20 | 0.7692 | 0.5214 | 0.7057 | 0.5128 |
| | $P_o = +1$ | | $P_o = -1$ | |

To study its effect, the membrane tension $\sigma$ is varied from 0.1 to 0.2, with $P_o$ fixed at either $+1$ or $-1$. Tightening the membrane; i.e. increasing $\sigma$, results in smaller membrane deformation. The largest values of the deformation and the locations where they occur are given in Table 1. Streamwise variations in the flow pressure and surface shear need not be demonstrated, as they follow the same patterns described above. It is, however, worth mentioning that the flow is, again, more sensitive to variations in $\sigma$, when $P_o$ is positive than when $P_o$ is negative.

# 6. CONCLUSION

In the present work, a mathematical model for the supersonic flow past a flat surface having a stretch of an elastic membrane with adjustable pressure is constructed, within the framework of the triple deck theory. The lower deck problem is supplemented with a membrane equation, so that the membrane shape can be determined simultaneously with the flow variables. A known lower deck solver is modified to handle this new feature efficiently. The numerical results agree with what one expects. As a hump (and much less so, as a dent), the membrane is found to lean downstream; being pushed by the flow. The convex (concave) shape of a hump (a dent) causes flow expansion (compression), which assists the membrane pressure in deforming the membrane further. However, interestingly enough, for the same magnitude of the membrane pressure $|P_o|$, a hump protrudes much more than a dent. The flow behaves as a hump (a dent) flow; experiencing separation when $|P_o|$ is sufficiently large.

# APPENDIX
## Cubic Splines for Smooth Membrane

The first sweep of every $k^{th}$ iteration produces new values of $F_{,XX}(i)$ for $i_B \leq i \leq i_E$. These values are used to determine new values of $F(i)$ for $i_B < i < i_E$, that are to be used in the next $k+1^{st}$ iteration.

To insure smoothness of the hump surface, we represent $F$ by cubic splines. For $i_B+1 \leq i \leq i_E$, we express $H = F_{,XX}$ over each interval $[i-1, i]$ of size $\Delta(i) = X(i) - X(i-1)$, by the linear expression

$$H = \frac{X - X(i)}{X(i-1) - X(i)} H(i-1) + \frac{X - X(i-1)}{X(i) - X(i-1)} H(i) \tag{A1}$$

By integrating (A1) the following relations for $F$ and $G = F_{,X}$ can be easily established

$$G(i) - G(i-1) = S(i) \tag{A2}$$
$$F(i) - F(i-1) = \Delta(i) G(i) - R(i) \tag{A3}$$

where $S(i)$ and $R(i)$ are calculated as follows

$$S(i) = \frac{\Delta(i)}{2} [H(i) + H(i-1)] \tag{A4}$$
$$R(i) = \frac{\Delta^2(i)}{6} [2H(i) + H(i-1)] \tag{A5}$$

We introduce the recurrence relation

$$F(i) = \overline{F}(i) - \widetilde{F}(i) G(i) \tag{A6}$$

Evaluating (A6) at $i-1$, substituting into (A3), and using (A2); then rearranging, we get

$$F(i) = [\overline{F}(i-1) + \widetilde{F}(i-1) S(i) - R(i)]$$
$$\qquad - \{\widetilde{F}(i-1) - \Delta(i)\} G(i) \tag{A7}$$

Comparison of (A7) to (A6) gives

$$\overline{F}(i) = \overline{F}(i-1) + \widetilde{F}(i-1) S(i) - R(i) \tag{A8}$$
$$\widetilde{F}(i) = \widetilde{F}(i-1) - \Delta(i) \tag{A9}$$

Starting with $\overline{F}(i_B) = 0$ and $\tilde{F}(i_B) = 0$, in order to satisfy the condition $F(i_B) = 0$; $\overline{F}(i)$ and $\tilde{F}(i)$ can be determined, recursively, for i= $i_B$+1→$i_E$. (A6) evaluated at i=$i_E$, with the condition $F(i_E) = 0$ invoked, gives

$$G(i_E) = \overline{F}(i_E) / \tilde{F}(i_E) \qquad (A10)$$

Recursive back-substitution, using (A2) then (A6), respectively gives $G(i)$ then $F(i)$, for i=$i_E$−1→$i_B$.